\documentclass{aastex63}

\usepackage{graphicx}
\usepackage{wrapfig}
\usepackage{natbib}

\newcommand{\be}{\begin{equation}}
\newcommand{\ee}{\end{equation}}
\newcommand{\nn}{\mbox{} \nonumber \\ \mbox{} }
\newcommand{\ba}{\begin{eqnarray}}
\newcommand{\ea}{\end{eqnarray}}
\newcommand{\om}{\omega}
\newcommand{\Alfven}{Alfv\'{e}n }

\newcommand\eg{{\it{e.g.\ }}}

\newcommand{\Bf}{{magnetic field}}

\newcommand{\NS}{neutron star}
\newcommand{\NSs}{{neutron stars}}
\newcommand{\EM}{electromagnetic}

\newcommand{\ms}{magnetosphere}
\newcommand{\mss}{magnetospheres}

\newcommand{\LC}{light cylinder}
\newcommand{\Lf}{Lorentz factor}

\begin{document}

\title{Brightness temperature constraints  on coherent processes  in magnetospheres of neutron stars}
\author{Maxim Lyutikov}
\affil{Department of Physics  and Astronomy, Purdue University,   525 Northwestern Avenue, West Lafayette, IN47907-2036, USA; lyutikov@purdue.edu}

%\author{
%Maxim Lyutikov\\
%Department of Physics and Astronomy, Purdue University, 525 Northwestern Avenue, West Lafayette, IN 47907-2036, USA   }

   \begin{abstract} 
   We discuss constraints that the observed brightness temperatures impose on coherent processes in pulsars and Fast Radio Bursts (FRBs), and in particular on the 
 hypothesis  of coherent curvature emission by bunches. 
We estimate the  peak brightness temperature that a bunch of charge $Ze$ can produce via synchrotron and/or curvature emission as $k_B T \sim (Z e)^2/\lambda$, where $\lambda$ is the typical emitted  wavelength.   
   We demonstrate that the  bunch's electrostatic energy required to produce observed brightness temperature is prohibitively high, of the order of the total  {\it  bulk } energy. We compare corresponding requirements for  the Free Electron Laser mechanism (Lyutikov 2021)  and find that  in that case the constraints are much easier satisfied.
   \end{abstract}

\section{Introduction}
Many modern day theories of coherent emission from pulsars/Fast Radio Burst (FRBs) accept the ``coherent curvature  emission by bunches" model,  formulated   in the early years of pulsar research \citep{1971ApJ...170..463G,1977ApJ...212..800C}, and go on calculating the details of this assumption. 
Extensive work in the 70s  through the day \citep{1977MNRAS.179..189B,1990MNRAS.247..529A,1992RSPTA.341..105M,1999ApJ...521..351M,2021MNRAS.500.4530M} demonstrated  that the model is not viable. 

 The  problems are many. For example, in the original  version of \cite{1971ApJ...170..463G}, the  bunching,   driven by weak radiation-reaction effect, is easily destroyed by minor velocity spread. Another problem is long times needed to create the bunches: in  a relativistically streaming plasma the processes in the beam frame are suppressed both by smaller rest-frame  beam/plasma density   if compared with the lab frame,  and relativistic freezing of any corresponding dynamical process  as viewed in the lab frame \citep[\eg][]{1999JPlPh..62...65L}.

Here we approach the question ``How justified is the assumption of the coherent emission by bunches'' from the observational side: let's assume that bunches are created, what is the corresponding conditions  to reproduce the observed properties? In particular,  creation of electrostatically repulsive bunches costs energy. - How much? Are the costs consistent with the model? (The answer is no.)

As a quantitative parameter that would measure the validity of the model we take the (awkwardly defined but universally used) quantity of brightness temperature.
\citep[Energetics is typically not an issue since radio carries minuscule amount of energy, though observations of Fast Radio Burst start to impose meaningful constraints on the plasma parameters at the source,][]{2019arXiv190103260L}.

 In case of pulsars and FRBs  the     brightness temperatures reach values in excess of  $10^{35}$ K \citep[\eg][]{1977puls.book.....M,Melrose00Review,2007Sci...318..777L}, and as high as  $\sim 10^{40}$ K in extreme cases \citep[\eg][]{2004ApJ...616..439S}. Can models of coherent emission by bunches reproduce those brightness temperature?
 
  We limit our approach to the class of ``antenna'' mechanism \citep[as opposed to plasma maser][]{1999ApJ...521..351M,1999ApJ...512..804L}.  That is:  all particles emit independently, but  due to the external driver they all emit in phase.  We also limit our consideration to models operating within the \mss\ of \NSs: simultaneous observations of radio and X-ray bursts \citep{2020Natur.587...54C,2021NatAs...5..372R,2020Natur.587...59B,2020ApJ...898L..29M}  unequivocally establishes  magnetospheric origin of FRBs,  as argued by \cite{2020arXiv200505093L}.

 Cyclotron and curvature emission are qualitatively very similar, but there is an important observational distinction: small cyclotron times typically cannot be resolved by the observing instruments, hence we see emission averaged over many  gyration  periods. Emitted and observed powers are equal then (if no bulk motion).  Curvature emission is different: particle emits once when its velocity is nearly  along the line of sight. At that  moment it is moving relativistically towards an observer, so that emitted and observed powers are different.

 %Second we calculate the energy price that needs to be paid to create charge bunches  (it is high) and test if models can satisfy this constraint. 

We illustrate the above  points with three examples: a more familiar synchrotron emission and a popular model of coherent curvature emission by bunches  \S \ref{synch},   and recently proposed Free Electron Laser mechanism, \S \ref{FEL}.

\section{Peak brightness temperatures of  synchrotron and curvature emission}

\subsection{Single particle}
\label{synch}

Let's start with the more familiar case of synchrotron emission. We are after   a subtle difference  between the emitted power and the observed one \citep[\eg][]{1968ApJ...151L.139S}.
The {\it average}  synchrotron power {\it emitted} by a particle (we stress the ``averaged'' and the ```emitted'' -   not ''instantaneously observed'')  is 
\be
P_{a,s} \approx \frac{e^2}{c} \gamma^2 \om_B^2
\label{Pas}
\ee
and typical frequency
\ba &&
\om_s \sim \gamma^2 \om_B
\nn &&
\om_B =\frac{e B}{m_e c}
\label{om1}
\ea
(For the sake of clarity we omit factors of unity.)

It can be derived from 
the relativistic Larmor formula for intensity of radiation 
   \be
   P_{a} = \frac{2}{3} e^2 \gamma^6 \left( a _\parallel ^2+ \frac{1}{\gamma^2}  a_\perp^2 \right),
   \label{I}
\ee
where $a _\parallel, a_\perp$ are acceleration along and perpendicular to the velocity. The quantity $P_a$ is the rate of energy loss by an electron seen in lab frame at the moment when the electron  has given velocity and acceleration, as measured in  the lab frame. 

For a particle on Larmor orbit with
\be
r_L =  \gamma \frac{c}{\om_B}
\ee
the  period of rotation and transverse acceleration are 
\ba&&
T_{rot} = \gamma \frac{2\pi}{\om_B}
\nn && 
a_\perp = \frac{c^2}{r_L} = \frac{c \om_B}{\gamma}
\ea
Hence  we derive relation (\ref{Pas}).

Relation  (\ref{I}) is the power {\it  emitted}  by an electron. It equals the {\it average } power seen by an observer stationary with respect to the gyration center, modulo angular factors of the order of unity - not dependent on the particle \Lf\ (in a sense  that an observer measuring energy flux through a given surface area, and projecting the result over the  whole  sky, would infer emitted power $\sim P_a$).  In other words, it's an integrated power seen by all observers spread out over $4 \pi$.

\begin{figure}[h!]
\centering
\includegraphics[width=.49\textwidth]{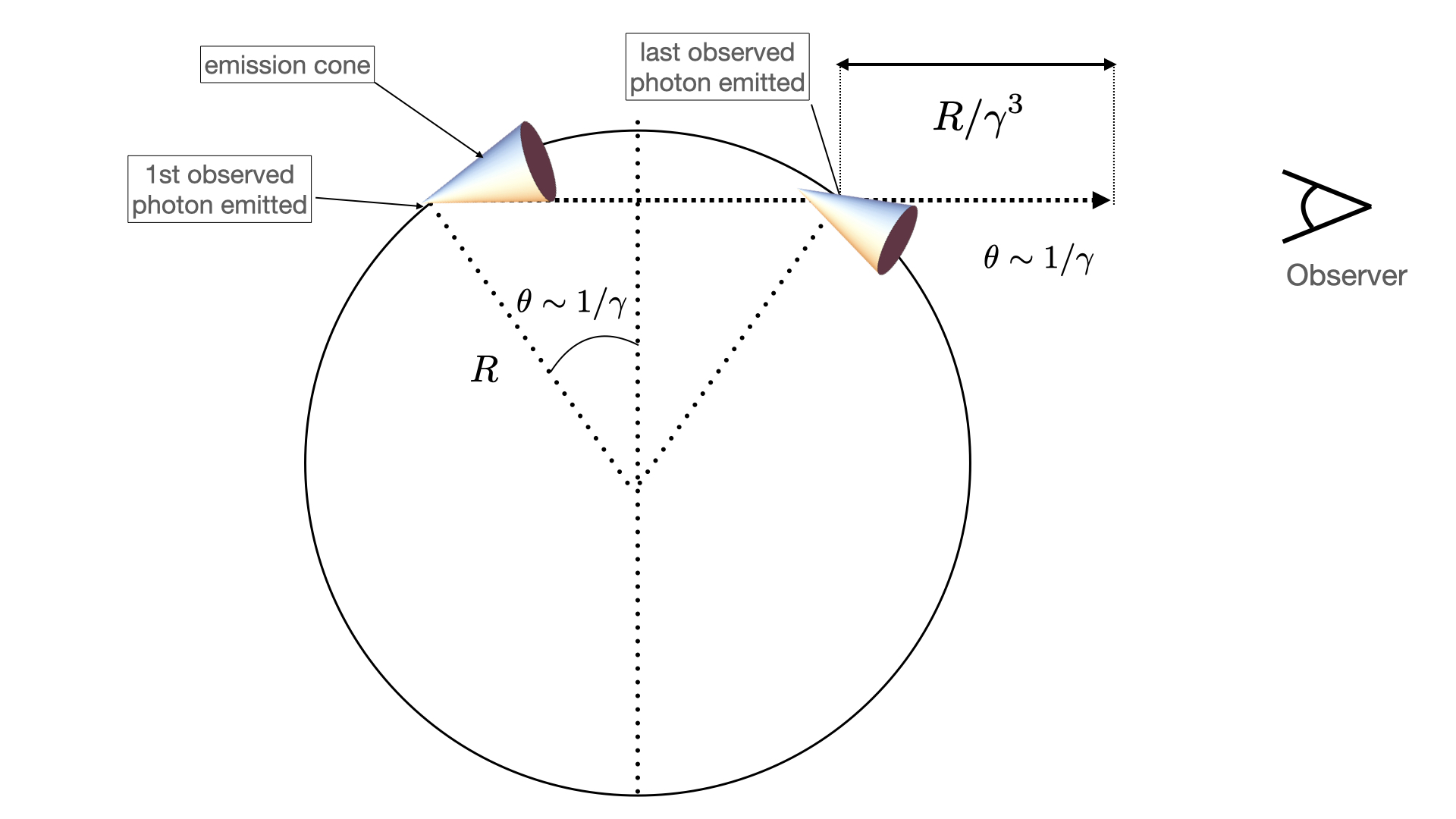} 
\includegraphics[width=.49\textwidth]{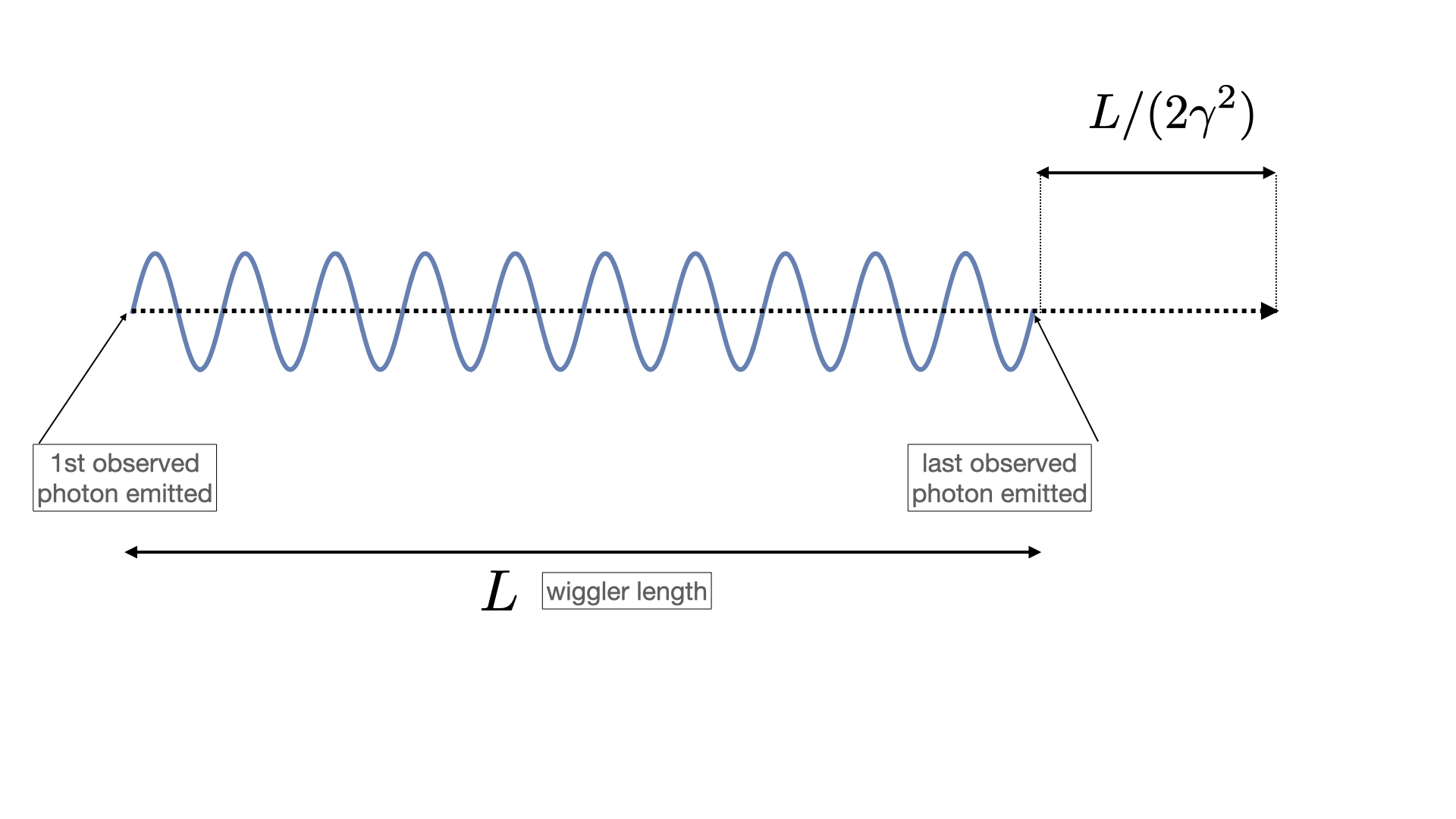}
\caption{Illustration of difference between emitted and observed powers for synchrotron/curvature emission when viewed on orbital time scale (left panel) and FEL emission (right panel).
In the case of synchrotron/curvature emission particle is moving clockwise  along the circle with radius $R$ and \Lf\ $\gamma$. It emits toward an observer for a time $\sim 2 (R/c) /\gamma$, but the first and last observed photons arrive within $\sim  2(R/c) /\gamma^3$. For $\gamma \gg 1$ the curvature of the trajectory is a higher order effect.
In the case of an FEL wiggler of length $L$ and particles propagating ``to the right'' all the emitted photons are concentrated with a region of $\sim L/(2 \gamma^2)$.   For \EM\ wiggler propagating ``to the left'' the interaction time and burst duration are further  reduced by $1+\beta \approx 2$.  }
\label{synchrotron-qualitative} 
\end{figure}

 For an observer in the gyration plane that average power comes in the form of bursts, when the direction of electron motion nearly coincides with the line of sight, see Fig. \ref{synchrotron-qualitative}

A relativistic  particle emits within a cone of $\sim 1/\gamma$. Hence a particle emits toward an observer for period
\be 
t_{em} \sim \frac{1}{\gamma} T_{rot} \sim  \frac{1}{\om_B}
\label{tem1}
\ee
Note that the radiation formation length
\be
l_r \sim \gamma^2 \lambda\sim c/\om_B \ll r_L
\ee is comparable to $c t_{em}$.

As the particle within a small angle $\sim 1/\gamma$ nearly catches-up with its radiation, the observed pulse duration
\be
t_{ob} \sim \frac{ t_{em} }{\gamma^2} =  \frac{2 \pi }{\gamma^2 \om_B}
\label{tob} 
\ee
Note that the inverse/Fourier transform of (\ref{tob}) gives the typical frequency (\ref{om1}). 

Now, energy emitted during time (\ref{tem1}) is
\be
E_{em} = \frac{e^2}{c} \gamma^2 \om_B
\ee
And the observed power
\be
P_{ob} = \frac{E_{em} }{t_{ob}} \sim  
\frac{e^2}{c} \gamma^4 \om_B^2
\ee

This is power during the peak emission in the middle of a short pulse.
The peak  power is $\gamma^2$ times larger than the average (\ref{Pas}). This is the difference between emitted and observed power. 
Qualitatively, an observer sees one  short bright burst per period of rotation. The peak brightness is much larger than the average one, by $\sim \gamma^2$.
In other words, all the energy we observe is emitted during $1/\gamma$ fraction of the orbit, but it arrives within time span $1/\gamma^2$ shorter.

Peak observed spectral power
\be
P_{ob,\om}= \frac{P_{ob}}{\om_s}  \sim  \frac{e^2}{c} \gamma^2 \om_B
\ee

Estimating  the  brightness temperature as
\be
k_B T_b \sim   P_{ob,\om} \frac{c}{(\om t_{ob})^2}
\ee
we find
\ba  &&
k_B T_b  \approx \frac{e^2}{\lambda}
\nn &&
\lambda = \frac{c}{\om}
\label{Tbs1}
\ea
Relation (\ref{Tbs1}) give the effective   brightness temperature during  a burst of synchrotron emission    for a single electrons producing synchrotron emission at wavelength $\lambda$.
Thus the brightness temperature of the emission peak of synchrotron and of the curvature emission is  approximately the electrostatic energy with the emitted wavelength.

Numerical estimate  gives
\be
T_b= 5 \times 10^{-5} \nu_9 \, {\rm K}
\ee
where $\nu_9 = \nu/10^9$ is the frequency of the emitted waves in GHz.

The  procedure outlined  above to estimate the peak brightness temperature   for synchrotron emission gives {\it the same}  brightness temperature for the curvature  emission as well, Eq. (\ref{Tbs1}). 
%\footnote{
Note that the observed  peak power of curvature emission is
\be
P_{ob, c} \sim  
\frac{e^2 }{c} \gamma^6 \left( \frac{c}{R_c}  \right)^2
\label{Pob}
\ee
is different from the emitted power by a factor $\gamma^2$. 
%}

\subsection{Coherent curvature emission by bunches}

The antenna mechanism is qualitatively based on the idea that 
if there is a bunch of $Z$ electrons that  has  a dimension smaller than a wave length $\leq \lambda$, it will produce an emission pulse with brightness temperature 
\be 
k_B T_b  \approx Z^2 \frac{e^2}{\lambda}.
\label{Tb1}
\ee
Scaling with $\propto Z^2$ indicates coherent process.
Hence  the number of electrons needed in a bunch to produce a given brightness temperature $T_b$ is 
\be
Z = \frac{\sqrt{k_B T_b \lambda}}{e}  = 2.5 \times 10^{16} T_{b,30}^{1/2} \lambda ^{1/2}
\ee
where we normalized brightness temperature to $10^{30}$ K, an appropriate scaling for pulsars and FRBs.

Importantly, relation (\ref{Tb1}) gives an estimate of the electrostatic Coulomb  energy $E_{Cmb,\, curv} $ required to create one bunch: 
\footnote{The corresponding equipartition \Bf, assuming that the bunch occupies volume $\sim\lambda^3$ is 
$$
B \sim \frac{ \sqrt{ k_B T_b}}{\lambda ^{3/2}} = 5 \times 10^7 \frac{T_{b,30} ^{1/2} }{ \lambda ^{3/2}}\, G
$$}.
\be
E_{Cmb,\, curv} \sim k_B T_b \sim 10^{14} T_{b,30}   {\rm erg}
\label{Ec}
\ee
The  ratio of electrostatic energy to create a bunch to the  bulk kinetic  energy  of the bunch $E_k \sim \gamma Z m_e c^2 $,
\be
\frac{E_{Cmb,\, curv} }{ E_k} \approx  \left(\frac{1}{\gamma} \frac{ (k_B T_b) }{m_e c^2}  \frac{ r_e  }{\lambda}   \right)^{1/2}  =
 \left(\gamma \frac{ (k_B T_b) }{m_e c^2}  \frac{ r_e  }{R_c}   \right)^{1/2}
 \label{22}
\ee
  Assuming  that $R_c \sim c/\Omega$, numerical estimates give
\be
 \frac{E_{Cmb,\, curv} }{ E_k}  = 7 \nu_9^{1/6} T_{b,30} ^{1/2},
 \label{Ecc} 
 \ee
 For Crab pulsar with period $0.03$ seconds, near the \LC.  For broader applications, the ratio $({ r_e  }/{R_c})^{1/2}$, Eq. (\ref{22}), microscopic to macroscopic parameters, does not vary much by changing the macroscopic one.
 
Relation (\ref{Ecc}) is  the main result: the electrostatic energy required to created charge bunches to produce coherent curvature emission of observed brightness temperature is prohibitively high, of the order of the total  {\it  bulk } energy.

Finally we note that the model of ``coherent emission by bunches" cannot be taken to the continuous limit (this would eliminate the extra $\gamma^2$ factor in the observed power). A constant flow of particles, even a charged one will  not produce any radiation. 

\section{FEL mechanism}
\label{FEL}
\begin{figure}[h!]
\centering
\includegraphics[width=.99\textwidth]{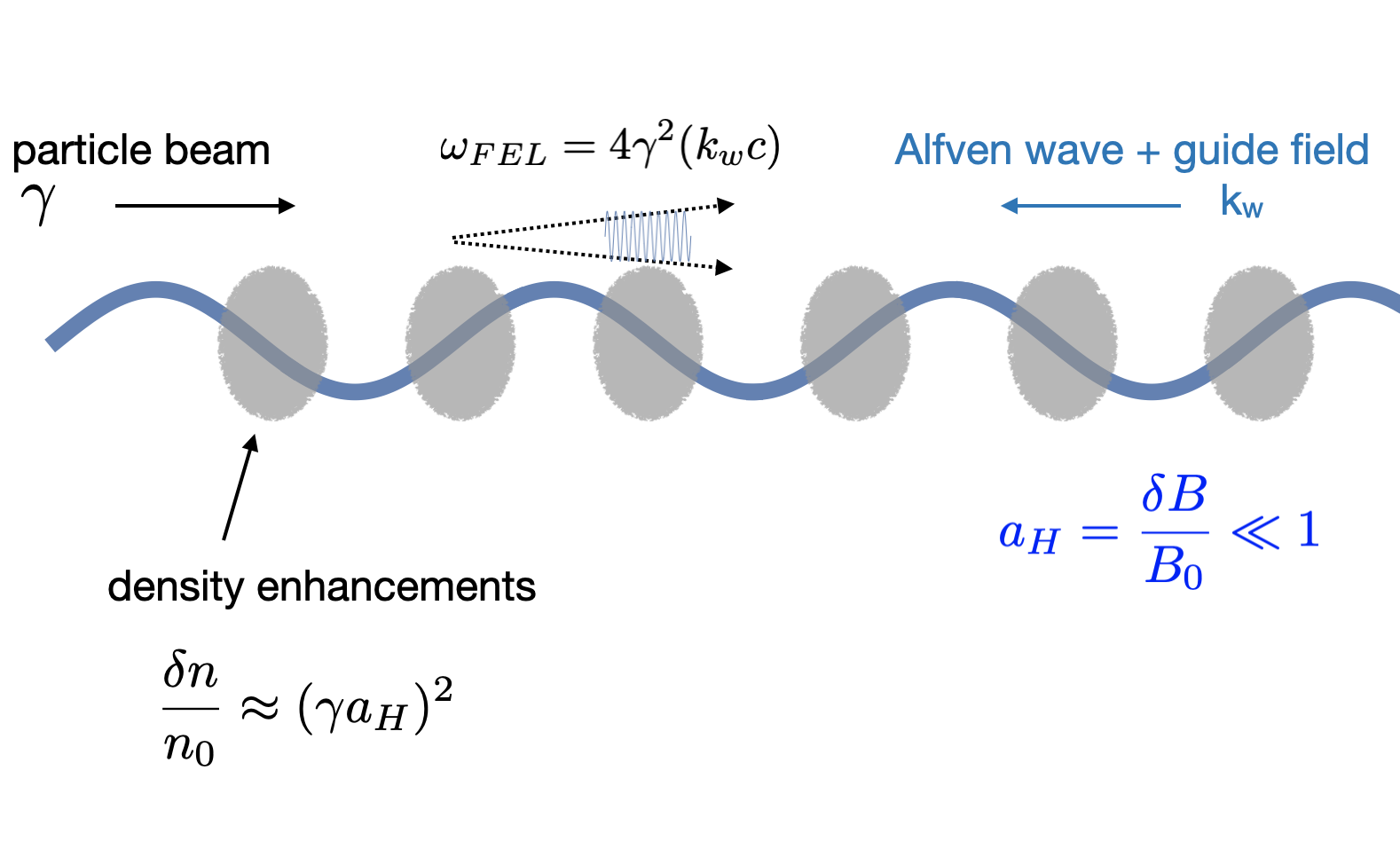} 
\caption{Cartoon of the FEL model of production of coherent emission.  An \Alfven wave  (a wiggler) with wavenumber $k_w$ is  propagating  to the left with relative amplitude $a_H$,  Eq. (\protect\ref{aH}). A beam of particles propagates to the right. In the guide field dominated regime particles move mostly  along the \Bf\ with velocity fluctuating at the double frequency; this creates density enhancements that  coherently Compton/Raman scatter the \Alfven wave.}
\label{FEL-Cartoon} 
\end{figure}

\cite{2021arXiv210207010L} developed a model of Free Electron Laser (FEL) for the production of coherent emission in pulsars (Crab in particular) and FRBs, Fig. \ref{FEL-Cartoon}. The model has many attractive features, including explanation for  some very  subtle observed relationships: (i) operates in a very broad range of \NS's parameters ({\it independent} of the value of the \Bf);
(ii) can tolerate mild  momentum spread of the beam $\Delta p /p \leq 1$; (iii)  reproduces (multiple) emission bands seen in Crab and FRBs; can also produce   broader   emission; (iv)  gives correct estimates for the brightness temperatures both in pulsars and FRBs; (v) explains correlation between polarization and spectral  properties (that narrow-band emission in FRBs correlated with linear polarization). 

The model assumes that \Alfven waves (\EM\ wigglers)  with relative amplitude 
\be
a_H = \frac{\delta B}{B_0} \ll 1
\label{aH} 
\ee
($\delta B$ is the fluctuating \Bf, $B_0$ is the guiding field)  and wave number $k_w$ propagates through \NS\ \ms. The wiggler shakes the electron beam. The resulting ponderomotive force, appearing due to the beat of the wiggler and the \EM\ wave amplified via parametric resonance,  leads to  the creation of charged bunches. The powerful wiggler then shakes the bunches producing coherent emission.

Single particle emitted power and frequency are
\ba &&
P_{a,FEL} \approx a_H^2  \frac{e^2}{c} \gamma^4  (k_w c)^2
\nn &&
\om_{FEL} \sim \gamma^2 (k_w c)
\label{PFEL}
\ea
Thus, the wiggler is Compton-scattered by the beam (FEL in guide field dominated regime cannot be treated as curvature emission in a wiggled field.)

The emitting particle continuously propagates towards an observer - this is an important different from the synchrotron and curvature emission (when observer sees a short burst).

Consider emission from particle propagating in a  static wiggler of length $L$ (relations are the same, within a factor of $\sim 2$,  for  the \EM\ wiggler propagating towards the particle), so that a particle emits for time $L/c$; but in observer frame this is shorter by $\gamma^2$. Total emitted energy is 
$\sim P_{a,FEL}  L/c$ and $t_{ob} \sim L/( \gamma^2 c)$.
The observed  power is thus  
\be
P_{ob,FEL} \approx a_H^2  \frac{e^2}{c} \gamma^6  (k_w c)^2
\label{PobFEL}
\ee
The 
peak spectral power
\be
P_\om = \frac{P_{ob,FEL}}{\om_{FEL} } =  a_H^2 e^2 \gamma^4 k_w = a_H^2 \frac{e^2}{c} \gamma^2 \om
\ee

The brightness temperature of the observed pulse is then 
\be
k_B T_b = a_H^2 \gamma^8 \frac{ e^2 c^2 k_w}{L^2 \om^2} =\left( \frac{ \gamma a_H }{\eta_w}  \right)^2 \frac{e^2}{\lambda} 
\ee
where we normalized the total wiggler length to the fluctuating wave number, 
$L = \eta_w /{k_w} $, $ \eta_ w \gg 1 $. (Factor $ { \gamma a_H }/ {\eta_w}$ is typically $\leq 1$).

The required bunching number $Z$ to produce given brightness temperature  is 
\be
Z  \approx \frac{\eta_w }{ \gamma a_H}  \frac{\sqrt{k_B T_b \lambda}}{e} 
\ee

The charged bunches in the FEL need not be static ones  - this involved huge energy as we discussed above.  The charged bunches can be dynamic (\eg, Langmuir oscillations driven by the wiggler for FEL operating in Raman regime), so here is no need to confine them electrostatically. Still, let us  take an extreme view and following the previous estimates demonstrate that FEL is consistent even with the  extreme case of statically produced charged bunches. 

The corresponding electrostatic energy 
\be
E_{Cmb,w} \sim Z^2 \frac{e^2}{\lambda} = \left(  \frac{\eta_w}{ \gamma a_H }\right)^2 k_B T_b
\ee
(we used $\lambda k_w \sim 1/\gamma^2$). In absolute values the quantity $E_{Cmb,w}$ is larger than $E_{Cmb,\, curv}$, (\ref{Ec}) since the factor in front of $k_B T_b$ is generally larger than unity.

Ratio of electrostatic energy 
over kinetic energy
\be
\frac{E_{Cmb,w} }{E_k} =   \frac{\eta_w }{a_H} \frac{ \sqrt{ k_B T_b} {k_w} \sqrt{ \lambda}}{m_e c^2}=
\left( \frac{\eta_w }{a_H}  \right) T_{30}^{1/2} \sqrt{\lambda}  \times 
\left\{ \begin{array}{cc}
2 \times 10^{-4} & \mbox{\rm Crab \LC}\\
3 \times 10^{-2} & \mbox{\rm near \NS}\\
\end{array}
\right.
\ee
(wavelength $\lambda$ is not normalized - as measured in centimeters.)
Smaller ratio of electrostatic to beam energy in this case can be traced to the fact that the frequency of  curvature emission scales as $\gamma^3$, while FEL as $\gamma^2$ - hence for similar scale $\sim$ \LC,  in the case of curvature emission $\gamma$ is smaller, hence smaller bulk energy.

Most importantly, in the case of the  FEL the bunching is driven, in some correctly implied sense, not  by the beam but by  the wiggler  (the axial bunching force $\propto \delta v \delta B$, while the  velocity drift component $ \delta v $ is also $\propto \delta B$). 

Ratio of electrostatic to wiggler energy within $\lambda^3$, $E_w \sim a_H^2 B_0^2 c \lambda ^3$,  evaluates to 
\be
\frac{E_{Cmb,w} }{E_w} = \frac{\eta_w^2 }{a_H^4}  \frac{ k_B T_b k_w}{B_0^2 \lambda^2}
=
\left( \frac{\eta_w^2 }{a_H^4}  \right) T_{30} {\lambda}^{-2}   \times 
\left\{ \begin{array}{cc}
 10^{-6} & \mbox{\rm at  Crab \LC}\\
10^{-19} b_q^{-2} & \mbox{\rm near magnetar}\\
\end{array}
\right.
\ee
where $ b_q= {B_0}/{B_Q}$, $B_Q =   m_e ^2 c^3 /({\hbar c})$. (Scaling to the quantum \Bf\  $B_Q$ is given since the model of  \cite{2021arXiv210207010L}  applies both to the inner regions of magnetars and the Crab's \LC.
Thus, mild wigglers can indeed create (have enough energy)  the required charged bunches.

 \section{Discussion}
 
 In this paper we argue that  a  popular simple  model of  ``coherent curvature  emission by bunches" is not self-consistent: the price in electrostatic energy that is needed to create charge bunches  is too high to explain the observed brightness temperatures.   This is our main conclusion.

Together with the well recognized problems of how to create charge bunches \citep{1977MNRAS.179..189B,1990MNRAS.247..529A,1992RSPTA.341..105M,1999ApJ...521..351M,2021MNRAS.500.4530M},   the unrealistic energetic requirements further assert  that coherent emission by bunches is not a viable pulsar/FRM emission mechanism.

  On  a more general issues, we    point out an   apparently  subtle confusion  in the literature between the emitted power and the observed one. The former can be measured in the electron rest frame, the latter depends on the location of the observer. Though synchrotron and curvature emission are qualitatively very similar,  the cyclotron motion is typically very fast, so that the use of the average power (\ref{Pas}) is, basically justified. If there is no bulk motion of plasma,  then the {\it average} observed power equals
{\it average} emitted power. This is not the case for curvature emission: particles stream along the field lines and appear in the line of sight once. At this moment the observed power is different from the emitted power by a factor $\sim \gamma ^2$. This is an important correction to the {\it observed}  power of curvature radiation.

% Since this issue is of the fundamental nature, let us discussed it in more details. Relation (\ref{I}) gives the power lost by a particle, as measured in the observer frame. {\it If} (!)  we had a collection of observers spread out over $4 \pi$, then adding their measurements of energy fluxes would amount to the power (\ref{I}). But our case is different: we do not have an access to the $4\pi$ measurements. An observer with the telescope's area $\ll 4\pi$  located along the particle's velocity measures a flux. Since the observer does not know  the angular distribution,  the observer  infers isotropic-equivalent power much larger than the particle emitted (for a multitude of observer spread out over $4 \pi$,  most would have reported  powers much smaller than the average).  Our observer, trying to estimate the properties of the curvature emission  typically scales it with $\gamma^4$, Eq. (\ref{om2}); instead, it should be scaled with $\gamma^6$, Eq. (\ref{Pob}).

  % The second point is that creation of electrostatically repulsing charges costs energy. As we argue, that cost is prohibitively high for  ``coherent curvature  emission by bunches".

 This work had been supported by 
NASA grants 80NSSC17K0757 and 80NSSC20K0910,   NSF grants 1903332 and  1908590.
I would like to thank Mikhail Medvedev and Alexandre Philippov  for discussions.

\bibliographystyle{apj}
\bibliography{/Users/maxim/Home/Research/BibTex}

\end{document}